\documentclass[aps,showpacs,preprintnumbers,amsmath,amssymb]{revtex4}
\UseRawInputEncoding
\usepackage[dvips]{epsfig}
\usepackage{graphicx}
\usepackage{amssymb}
\usepackage{color}
\usepackage{enumerate}

\begin{document}
\baselineskip=0.8 cm
\title{\bf Polarized image of a Schwarzschild black hole with a thin accretion disk as photon couples to Weyl tensor}

\author{Zelin Zhang$^{1}$,
Songbai Chen$^{1,2}$\footnote{Corresponding author: csb3752@hunnu.edu.cn}, Xin Qin$^{1}$,
Jiliang Jing$^{1,2}$ \footnote{jljing@hunnu.edu.cn}}
\affiliation{ $ ^1$ Department of Physics, Key Laboratory of Low Dimensional Quantum Structures
and Quantum Control of Ministry of Education, Synergetic Innovation Center for Quantum Effects and Applications, Hunan
Normal University,  Changsha, Hunan 410081, People's Republic of China
\\
$ ^2$Center for Gravitation and Cosmology, College of Physical Science and Technology, Yangzhou University, Yangzhou 225009, People's Republic of China}

\begin{abstract}
\baselineskip=0.6 cm
\begin{center}
{\bf Abstract}
\end{center}

We have studied polarized image of a Schwarzschild black hole with an equatorial thin accretion disk as photon couples to Weyl tensor. The birefringence of photon originating from the coupling affect the black hole shadow, the thin disk pattern and its luminosity distribution. We also analyze the observed polarized intensity in the sky plane.  The observed polarized intensity in the bright region is stronger than that in the darker region. The stronger effect of the coupling on the observed polarized vector appears only in the bright region close to black hole. These features in the polarized image could help us to understand black hole shadow, the thin accretion disk and the coupling between photon and Weyl tensor.

\end{abstract}

 \maketitle
\newpage
\section{Introduction}

The detection of gravitational waves \cite{P1,P2,P3,P4,P5} together with the release of the first images of the black hole M87* \cite{fbhs1,fbhs6} indicates that the observational black hole astronomy has been entered an exciting era of rapid progress. Recently, the Event Horizon Telescope (EHT) collaboration have also released the first polarized images of the black hole M87* \cite{poimag1,poimag2,poimag3}. The brightness of the surrounding
emission region and the corresponding polarization pattern provide a wealth of information about the electromagnetic emissions from the black hole's vicinity,  which is helpful to understand the material distribution, the electromagnetic interaction and  the accretion process near the black hole. This has encouraged a lot of effort to make theoretical research on polarized images of black holes because it could help us to put insight into physics in the strong field region near black holes by comparing the theoretical polarized patterns with the observed polarization signatures \cite{poth1,poth2,poth3}. Recently, the polarized images of axisymmetric fluid orbiting in various magnetic field has been investigated for a Kerr black hole with a simple model \cite{pokerr}. It is shown that the magnetic field configuration, together with black hole spin and observer inclination, affects the polarization signatures of the image including photon ring.

In general, the  distribution of polarized intensity and polarized direction in the  black hole's image depend on
the propagation of polarized light in the spacetime, which is determined by the parameters of background black hole, the dynamical properties of photon itself and the interactions between photon and other fields. It is well known that electromagnetic force and gravity are two kinds of fundamental forces in nature and then the interaction between the electromagnetic and gravitational fields should be important in physics. In the standard Einstein-Maxwell theory, there is only a quadratic
term of Maxwell tensor related directly to electromagnetic field, which
can also be understood as an interaction between Maxwell field and the metric tensor.
However, the interactions between electromagnetic field and curvature tensor are not included
in this theory.  Actually, in a curved background spacetime,  such kind of the couplings with curvature tensors
could be appeared naturally in quantum electrodynamics with the photon effective action
originating from one-loop vacuum polarization \cite{Drummond}.  Although these curvature tensor corrections appear firstly as an effective description of quantum effects, they may also occur near classical compact
astrophysical objects with high mass density and a strong gravitational field around the supermassive black holes at
the center of galaxies \cite{Dereli}. And then the models with arbitrary coupling constant have been investigated widely
for some physical motivation \cite{weyl1,weyl2,weyl3,weyl4,weyl5,weyl6,weyl7,weyl8,weyl9,weyl10,weyl11,weyl12,weyl13,weyl14,weyl15,weyl16}.

In this paper, we focus on a simple interaction model where Maxwell field couples to Weyl tensor. The main reason is that Weyl tensor is an important tensor in general relativity since it describes a type of gravitational distortion in the spacetime.
The coupling between Maxwell field and Weyl tensor changes both the path and the maximum velocity of photon propagation, which could result in  the ``superluminal" phenomenon \cite{Drummond,Caip,Cho1,Lorenci}.
The optical behaviors of the coupled photons in the extended Weyl correction model have been studied in the strong field region \cite{sb0,xy}, which shows that the measurement of relativistic images and time delay in the strong field can provide a mechanism to detect the polarization direction of coupled photons. The double shadow of a regular phantom black hole has been studied due to birefringence originating from the coupling between Maxwell field and Weyl tensor \cite{sb2}. However, it is still unclear what effects of the coupling between photon and Weyl tensor on the polarized image of a black hole with a thin accretion disk. The motivation in this paper is to study the image of a thin accretion disk around a black hole under the interaction between Maxwell field and Weyl tensor,  and then probe the effect of the coupling on the corresponding polarized patterns.

The paper is organized as follows: In section II,  we present equation of motion for the photons coupled to Weyl
tensor in a Schwarzschild black hole spacetime. In section III, we investigate the polarized image of a Schwarzschild black hole with a thin disk by polarized light coupled with Weyl tensor.  Finally, we end the paper with a summary.

\section{Equation of motion for the photons coupled to Weyl tensor}

Let us now to review briefly equation of motion for the photons coupled to Weyl tensor. In a curved spacetime, the action of the electromagnetic field coupled to Weyl tensor can be expressed as
\begin{eqnarray}
S=\int d^4x\sqrt{-g}\bigg[\frac{R}{16\pi G}-\frac{1}{4}\bigg(F_{\mu\nu}F^{\mu\nu}-4\alpha
C^{\mu\nu\rho\sigma}F_{\mu\nu}F_{\rho\sigma}\bigg)\bigg].\label{acts}
\end{eqnarray}
Here $F_{\mu\nu}$ and $\alpha$ are the usual electromagnetic tensor and the coupling constant with dimension of length-squared, respectively. Weyl tensor $C_{\mu\nu\rho\sigma}$ is defined by
\begin{eqnarray}
C_{\mu\nu\rho\sigma}=R_{\mu\nu\rho\sigma}-(g_{\mu[\rho}R_{\sigma]\nu}-g_{\nu[\rho}R_{\sigma]\mu})+\frac{1}{3}R
g_{\mu[\rho}g_{\sigma]\nu},
\end{eqnarray}
where the brackets around indices refers to the antisymmetric part. Varying the
action (\ref{acts}) with respect to electromagnetic vector $A_{\mu}$, one can obtain the corrected Maxwell equation
\begin{eqnarray}
\nabla_{\mu}\bigg(F^{\mu\nu}-4\alpha
C^{\mu\nu\rho\sigma}F_{\rho\sigma}\bigg)=0.\label{WE}
\end{eqnarray}
In order to obtain the equation of motion of the coupled photons from the above corrected Maxwell equation (\ref{WE}), one can adopt the short wave approximation where the wavelength $\lambda$ of photon is much smaller than a typical curvature scale $L$, but is larger than the electron Compton wavelength $\lambda_c$. In this approximation, the electromagnetic tensor can be written as a simple form
\begin{eqnarray}
F_{\mu\nu}=f_{\mu\nu}e^{i\theta},\label{ef1}
\end{eqnarray}
with a slowly varying amplitude $f_{\mu\nu}$ and a rapidly varying phase $\theta$.  And then the derivative term $f_{\mu\nu;\lambda}$ can be neglected because the amplitude $f_{\mu\nu}$ is slowly varying \cite{Drummond,Caip,Cho1,Lorenci}.  The wave vector $k_{\mu}=\partial_{\mu}\theta$ can be treated as the coupled photon momentum as in the usual theory of particle. From the Bianchi identity,
one can find that the amplitude $f_{\mu\nu}$ has a form $f_{\mu\nu}=k_{\mu}a_{\nu}-k_{\nu}a_{\mu}$,
where $a_{\mu}$ is the polarization vector satisfying the condition that
$k_{\mu}a^{\mu}=0$. Combining Eq.(\ref{ef1}) with Eq.(\ref{WE}), one can find that the equation of motion of photon coupling to Weyl tensor becomes
\begin{eqnarray}
k_{\mu}k^{\mu}a^{\nu}+8\alpha
C^{\mu\nu\rho\sigma}k_{\sigma}k_{\mu}a_{\rho}=0.\label{WE2}
\end{eqnarray}
Obviously, the coupling with Weyl tensor changes the propagation of the coupled photon
in the background spacetime.

For a Schwarzschild black hole spacetime, one can introduce the vierbein fields
\begin{eqnarray}
e^a_{\mu}=diag(\sqrt{f},\;\frac{1}{\sqrt{f}},\;r,\;r\sin\theta),
\end{eqnarray}
and rewrite the black hole metric as
$g_{\mu\nu}=\eta_{ab}e^a_{\mu}e^b_{\nu}$, where $\eta_{ab}$ is the Minkowski metric and $f=1-\frac{2M}{r}$. With the antisymmetric combination of
vierbeins defined in \cite{Drummond}
\begin{eqnarray}
U^{ab}_{\mu\nu}=e^a_{\mu}e^b_{\nu}-e^a_{\nu}e^b_{\mu},
\end{eqnarray}
Weyl tensor can be further simplified as
\begin{eqnarray}
C_{\mu\nu\rho\sigma}&=&\mathcal{A}\bigg(2U^{01}_{\mu\nu}U^{01}_{\rho\sigma}-
U^{02}_{\mu\nu}U^{02}_{\rho\sigma}-U^{03}_{\mu\nu}U^{03}_{\rho\sigma}
+U^{12}_{\mu\nu}U^{12}_{\rho\sigma}+U^{13}_{\mu\nu}U^{13}_{\rho\sigma}-
2U^{23}_{\mu\nu}U^{23}_{\rho\sigma}\bigg),
\end{eqnarray}
with
\begin{eqnarray}
\mathcal{A}=-\frac{M}{r^3}.
\end{eqnarray}
Introducing three linear combinations of momentum components \cite{Drummond}
\begin{eqnarray}
l_{\nu}=k^{\mu}U^{01}_{\mu\nu},\;\;\;\;
n_{\nu}=k^{\mu}U^{02}_{\mu\nu},\;\;\;\;
m_{\nu}=k^{\mu}U^{23}_{\mu\nu},
\end{eqnarray}
together with the dependent combinations
\begin{eqnarray}
&&p_{\nu}=k^{\mu}U^{12}_{\mu\nu}=\frac{1}{k^0}\bigg(k^1n_{\nu}-k^2l_{\nu}\bigg),\nonumber\\
&&r_{\nu}=k^{\mu}U^{03}_{\mu\nu}=\frac{1}{k^2}\bigg(k^0m_{\nu}+k^3l_{\nu}\bigg),\nonumber\\
&&q_{\nu}=k^{\mu}U^{13}_{\mu\nu}=\frac{k^1}{k^0}m_{\nu}+
\frac{k^1k^3}{k^2k^0}n_{\nu}-\frac{k^3}{k^0}l_{\nu},\label{vect3}
\end{eqnarray}
the equation of motion of the coupled photon (\ref{WE2}) can be simplified further as a set of equations for three independent polarisation components $a\cdot l$, $a\cdot n$, and $a\cdot m$,
\begin{eqnarray}
\bigg(\begin{array}{ccc}
K_{11}&0&0\\
K_{21}&K_{22}&
K_{23}\\
0&0&K_{33}
\end{array}\bigg)
\bigg(\begin{array}{c}
a \cdot l\\
a \cdot n
\\
a \cdot m
\end{array}\bigg)=0.\label{Kk}
\end{eqnarray}
Here we do not list the coefficients $K_{ij}$ ( for more details see refs. \cite{Drummond,sb0,sb2} and reference therein). As in refs.\cite{Drummond,sb0,sb2,xy}, there are only two physical solutions for Eq.(\ref{Kk}). The first solution is
\begin{eqnarray}
&&(1+16\alpha \mathcal{A})(g^{00}k_0k_0+g^{11}k_1k_1)+(1-8\alpha \mathcal{A})(g^{22}k_2k_2+g^{33}k_3k_3)=0, \label{Kk31}
\end{eqnarray}
which corresponds to the case the polarization vector $a_{\mu}$ is proportional to $l_{\mu}$.  The second one is
\begin{eqnarray}
&&(1-8\alpha \mathcal{A})(g_{00}k^0k^0+g_{11}k^1k^1)+(1+16\alpha \mathcal{A})(g_{22}k^2k^2+g_{33}k^3k^3)=0,\label{Kk32}
\end{eqnarray}
which means that the polarization vector $a_{\mu}=\lambda m_{\mu}$. From above two equations, it is easy to find that
the equation of motion is different for the photon with different polarizations, which leads to a phenomenon of birefringence of photon and then it can be expect that the coupling with Weyl tensor will affect the image of black hole with a thin disk and its luminosity. The light cone conditions (\ref{Kk31}) and (\ref{Kk32})
imply that the motion of the coupled photons is geodesic in the effective metric $\gamma_{\mu\nu}$ rather than in
the original metric $g_{\mu\nu}$ \cite{effmetr}.  The effective metric for the
coupled photon can be expressed as \cite{sb0}
\begin{eqnarray}
ds^2&=&-A(r)dt^2+B(r)dr^2+C(r)W(r)^{-1}(d\theta^2+\sin^2\theta d\phi^2),\label{l1}
\end{eqnarray}
where $A(r)=B(r)^{-1}=1-\frac{2M}{r}$ and $C(r)=r^2$. The quantity $W(r)$ is
\begin{eqnarray}
W(r)=\frac{r^3-8\alpha M}{r^3+16\alpha M},\label{v111}
\end{eqnarray}
for photon with the polarization along $l_{\mu}$ (PPL ) and is
\begin{eqnarray}
W(r)=\frac{r^3+16\alpha M}{r^3-8\alpha M},\label{v112}
\end{eqnarray}
for photon with the polarization along $m_{\mu}$ (PPM ). With the increase of the coupling parameter $\alpha$, one can find that the inner circular orbit radius $r_{ph}$ increases for PPL and decreases for PPM \cite{sb0}.
\begin{figure}
\includegraphics[width=16cm]{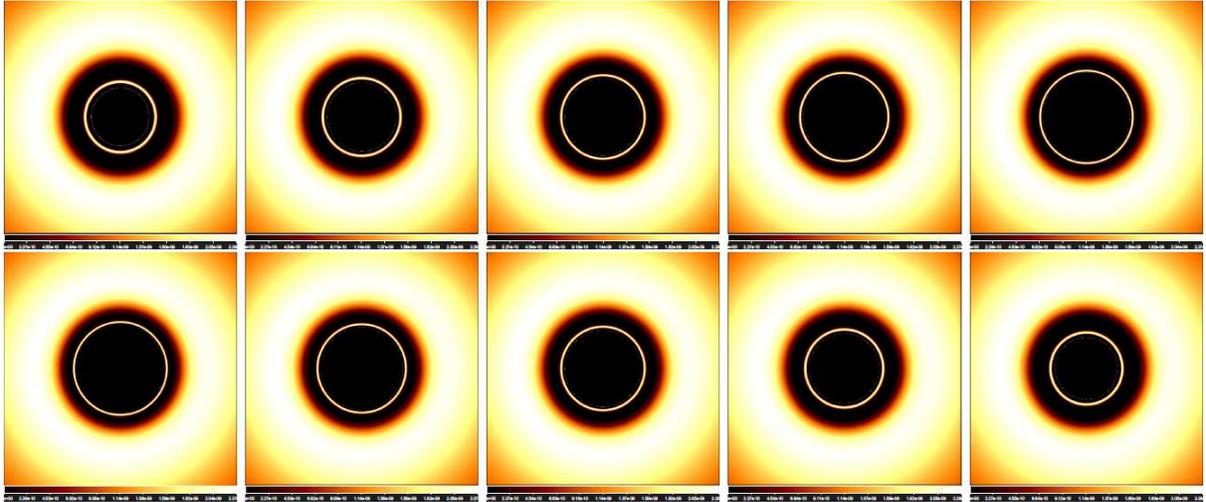}
\caption{Image of a geometrically thin infinite accretion disk around a Schwarzschild
black hole by polarized light coupled with Weyl tensor for the observer with $r_o=100M$ and $\theta_o=0^{\circ}$. The upper row is for the PPL and the bottle row is for the PPM. In each row, the coupling parameter $\alpha$ from left to right is taken to $-0.4$, $-0.2$, $0$, $0.2$ and $0.4$, respectively. Here, we set $M=1$.}
\label{f1}
\end{figure}

\section{Image of a Schwarzschild black hole with a thin disk by polarized light coupled with Weyl tensor}

\begin{figure}
\includegraphics[width=16cm]{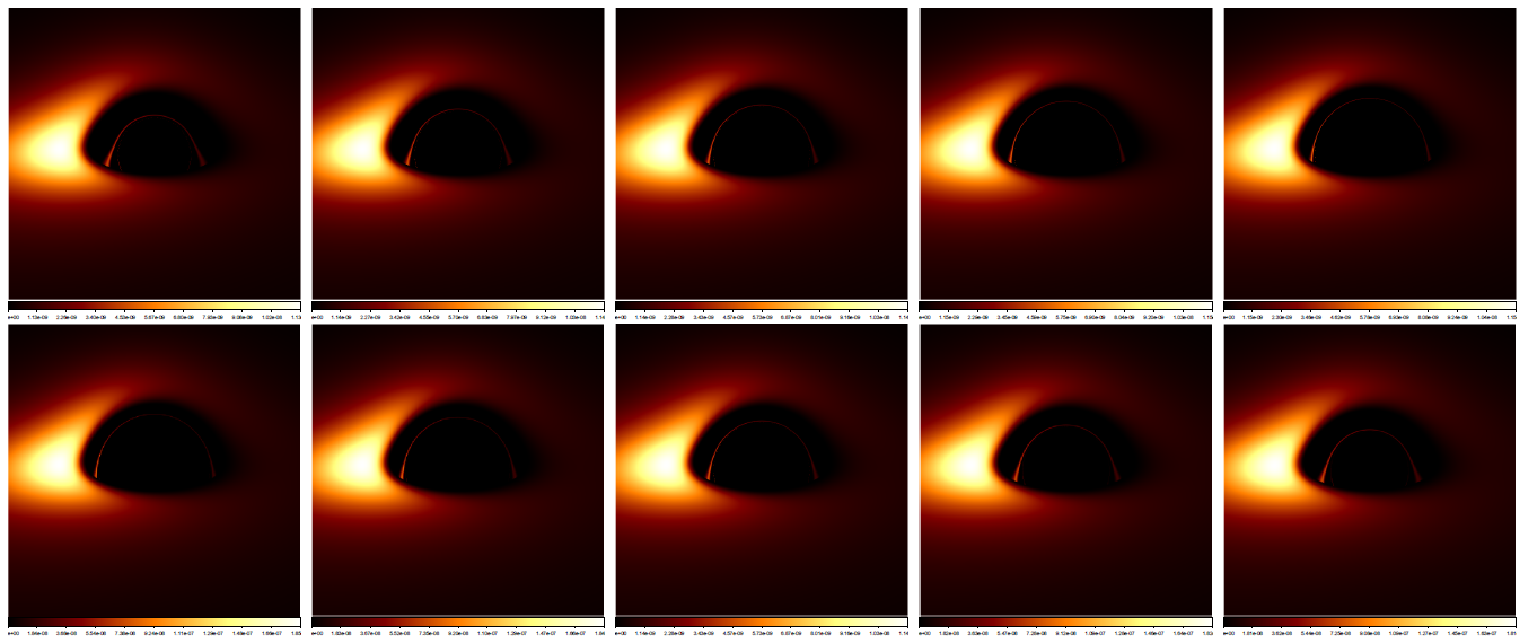}
\caption{Image of a geometrically thin infinite accretion disk around a Schwarzschild
black hole by polarized light coupled with Weyl tensor for the observer with $r_o=100M$ and $\theta_o=70^{\circ}$. The upper row is for the PPL and the bottle row is for the PPM. In each row, the coupling parameter $\alpha$ from left to right is taken to $-0.4$, $-0.2$, $0$, $0.2$ and $0.4$, respectively. Here, we set $M=1$.}
\label{f2}
\end{figure}

\begin{figure}
\includegraphics[width=16cm]{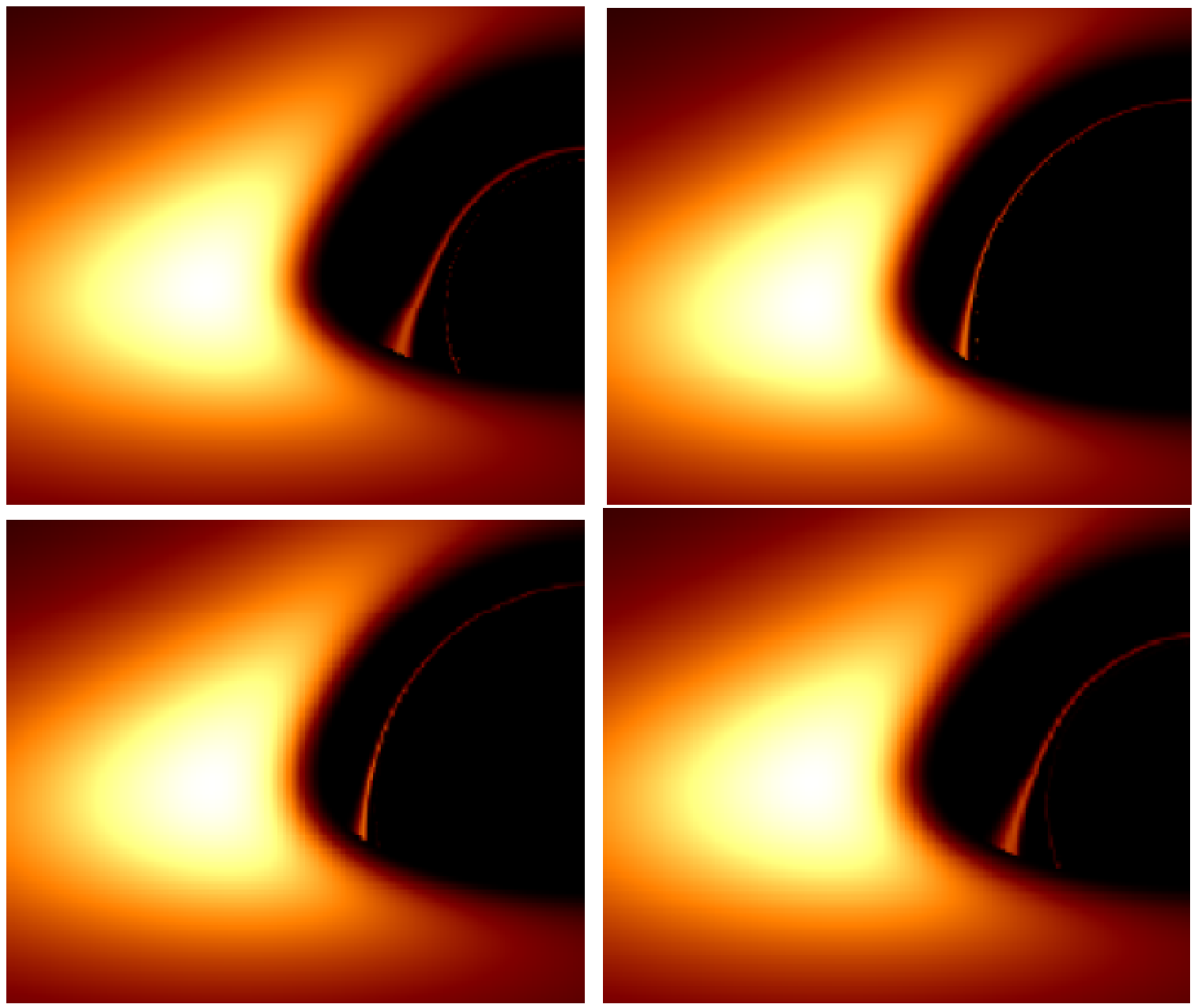}
\caption{The partial enlargement of images in Fig.(\ref{f2}). The upper row is for the PPL and the bottle row is for the PPM. In each row, the left panel and the right panel correspond to the case with $\alpha=-0.4$ and $\alpha=0.4$, respectively. Here, we set $M=1$.}
\label{f3}
\end{figure}

\begin{figure}
\includegraphics[width=4.0cm]{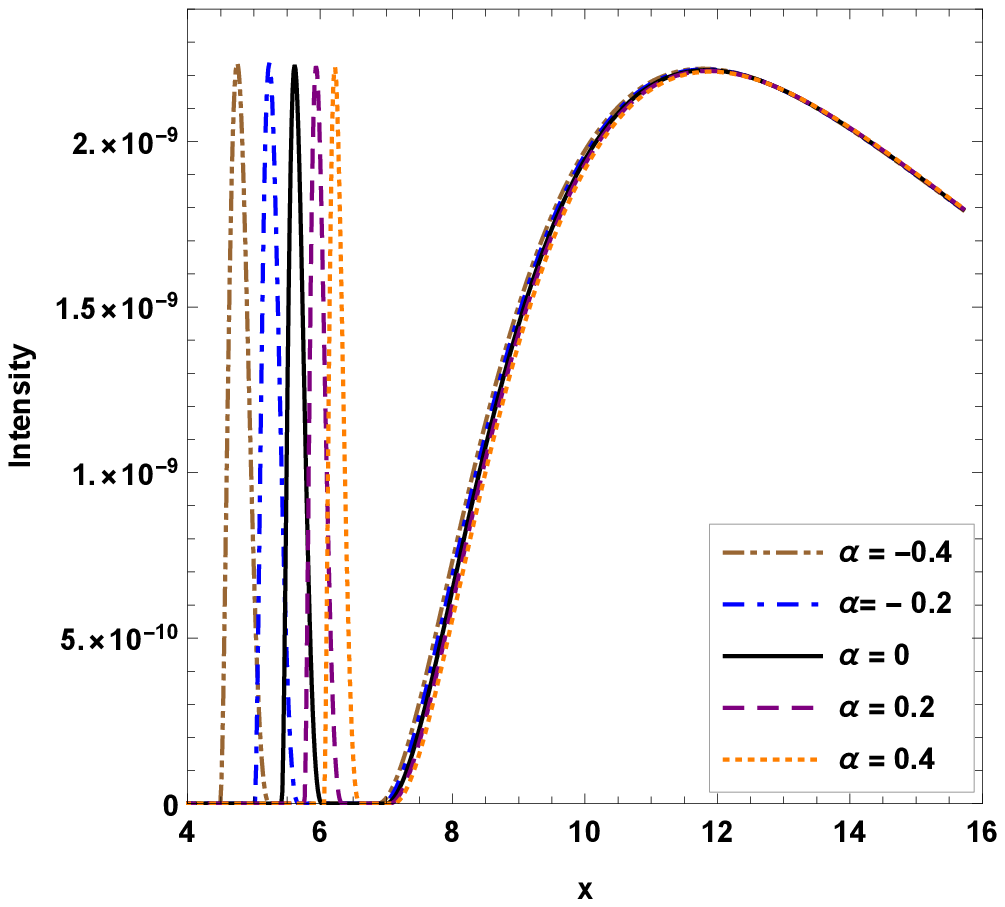}
\includegraphics[width=4.0cm]{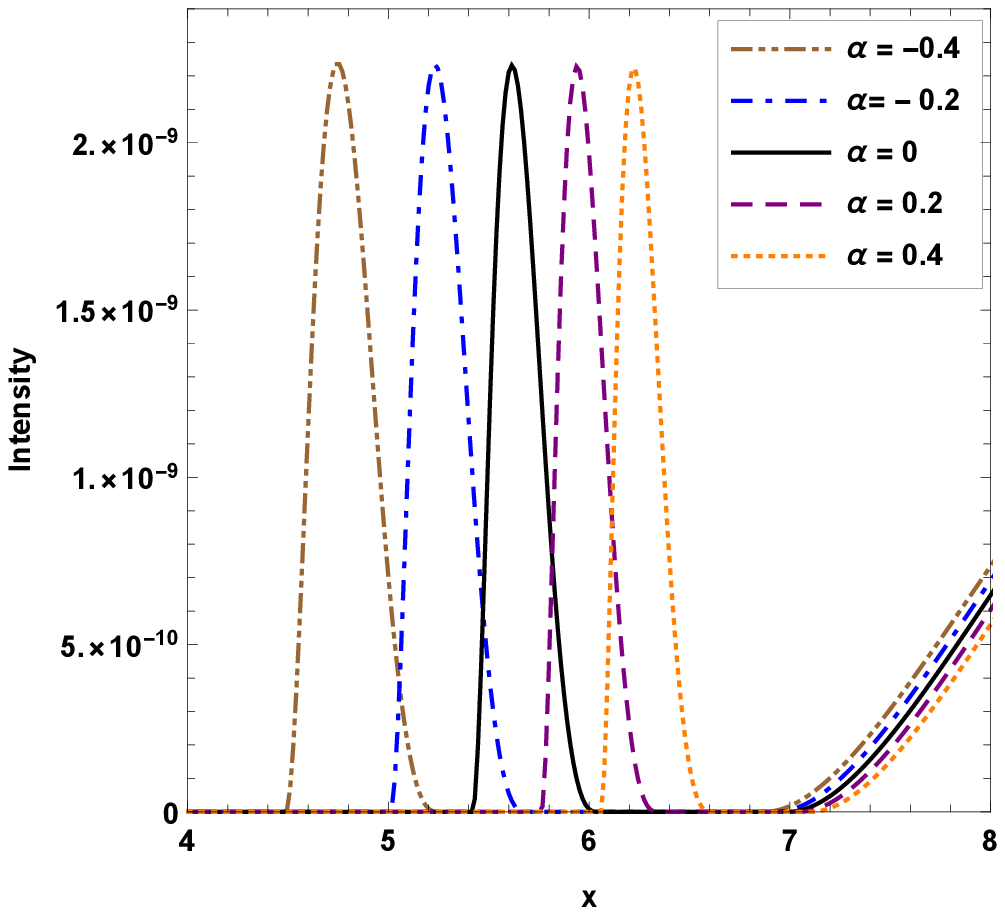}
\includegraphics[width=4.0cm]{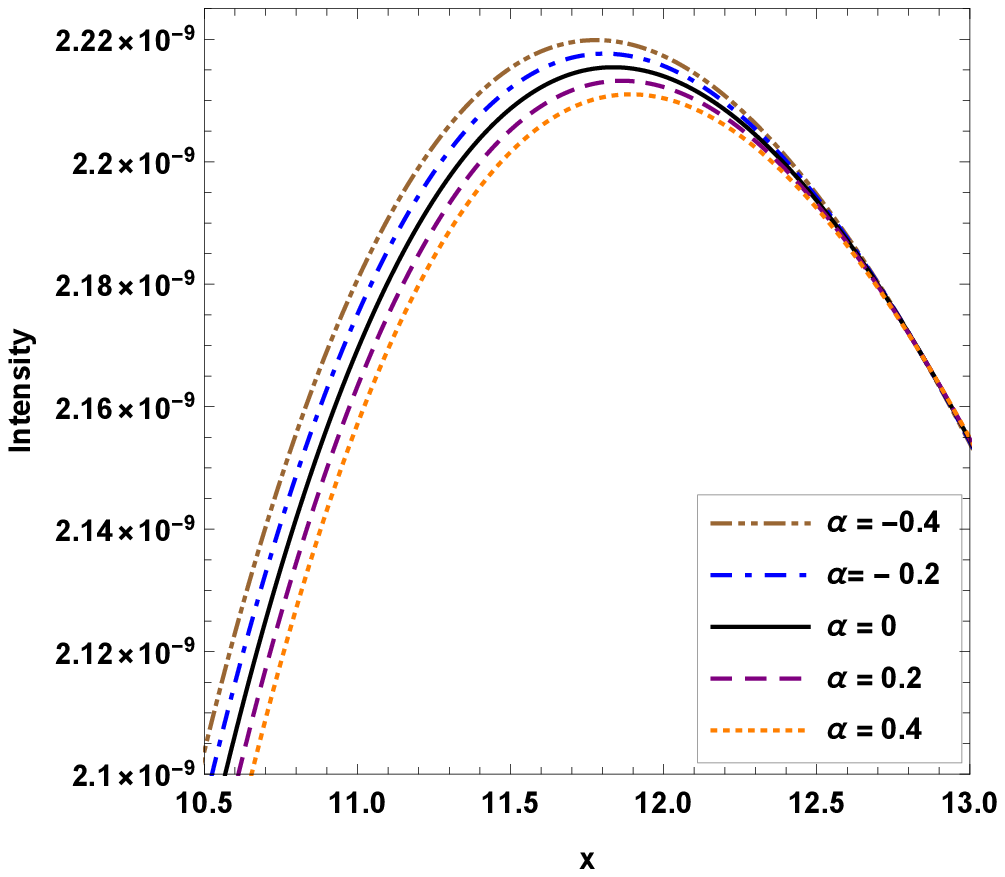}
\includegraphics[width=4.0cm]{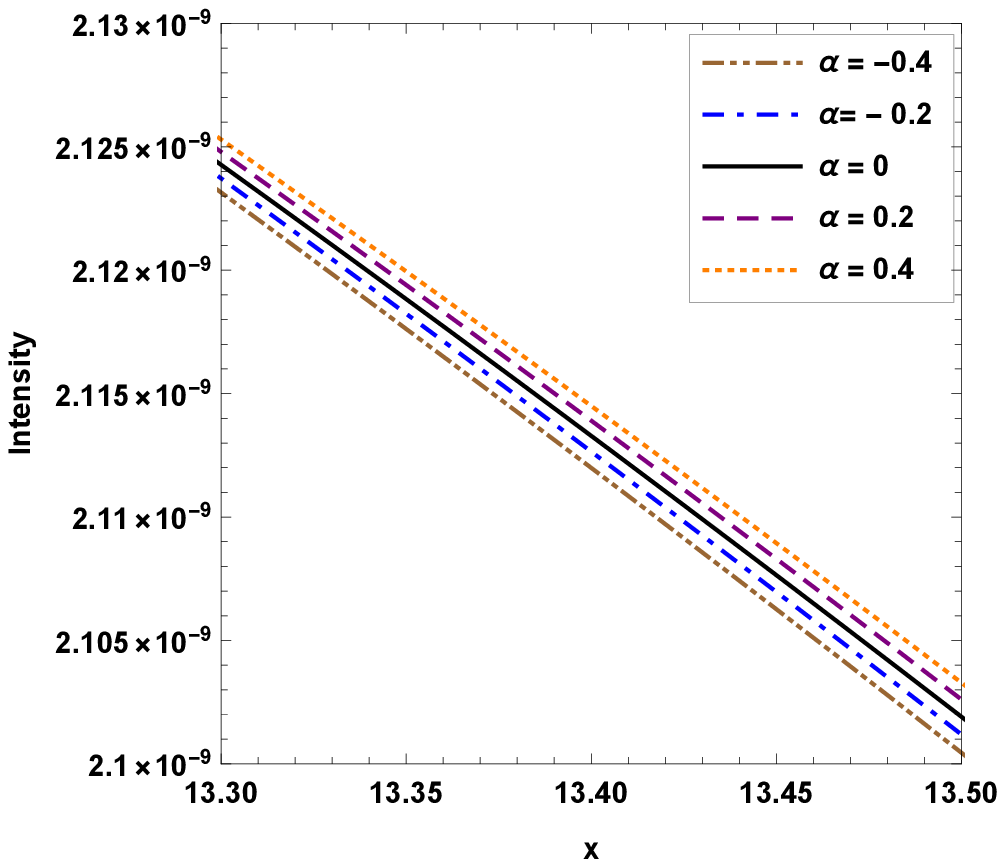}\\
\includegraphics[width=4.0cm]{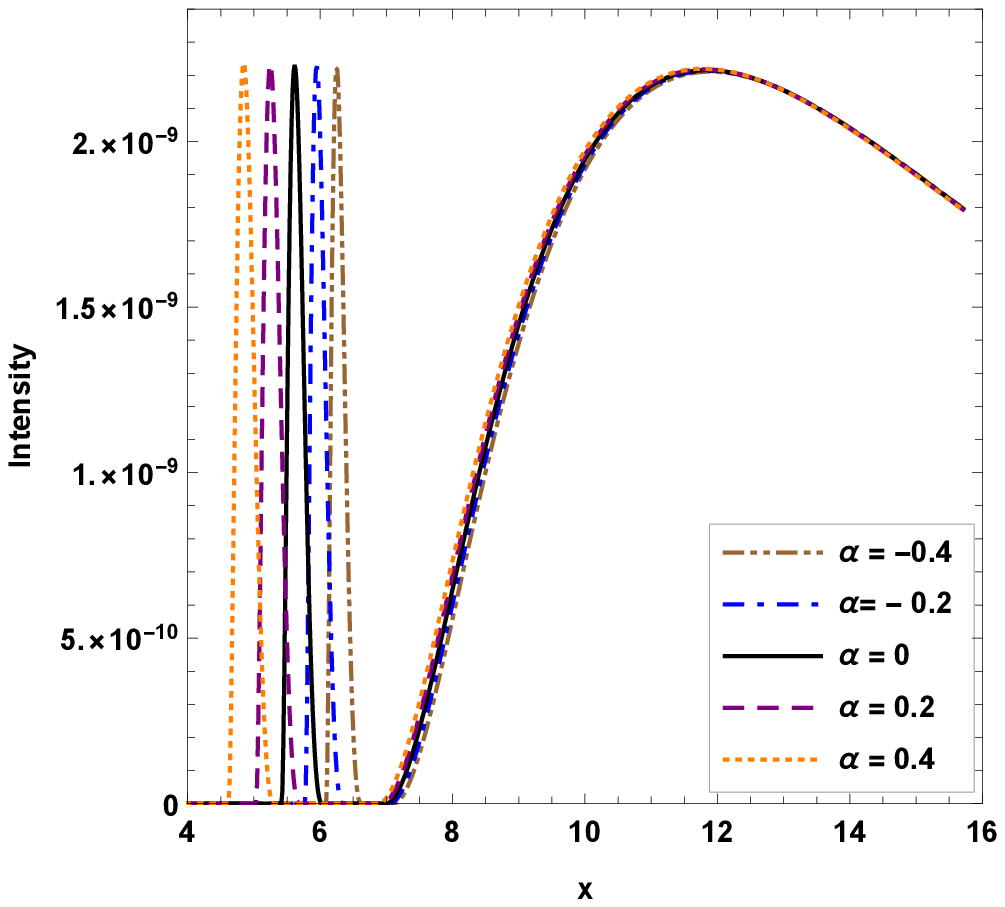}
\includegraphics[width=4.0cm]{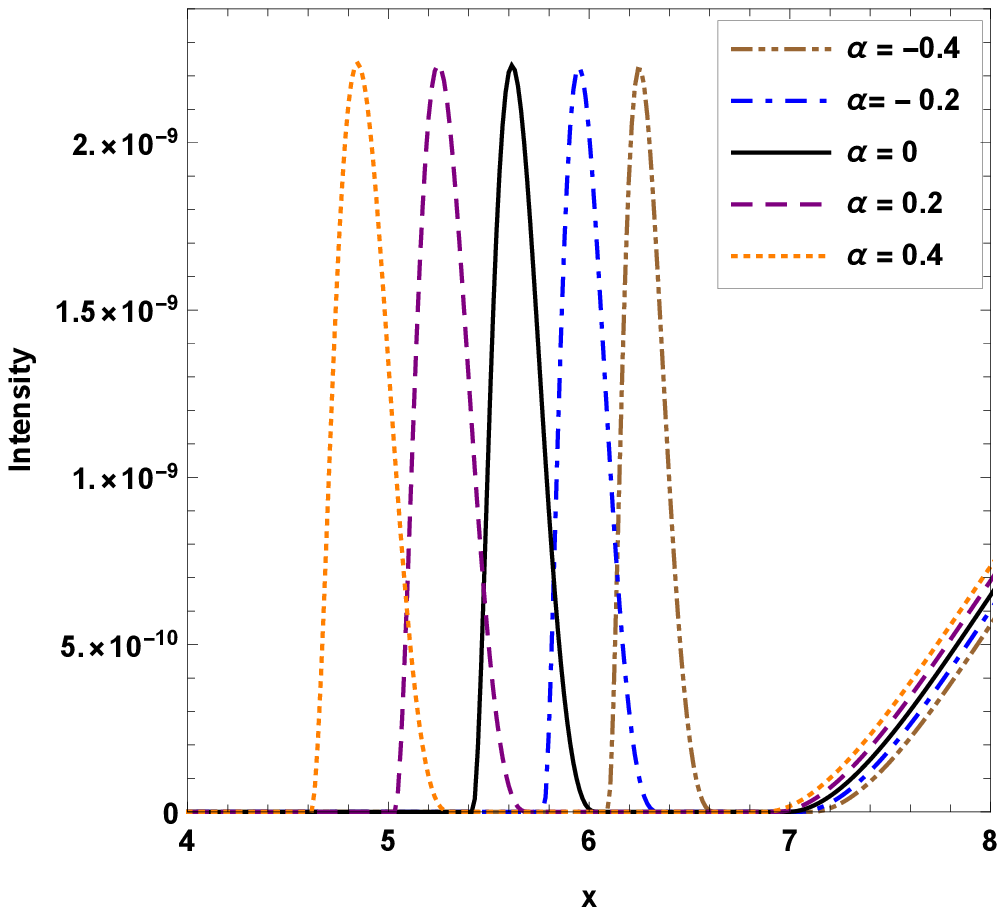}
\includegraphics[width=4.0cm]{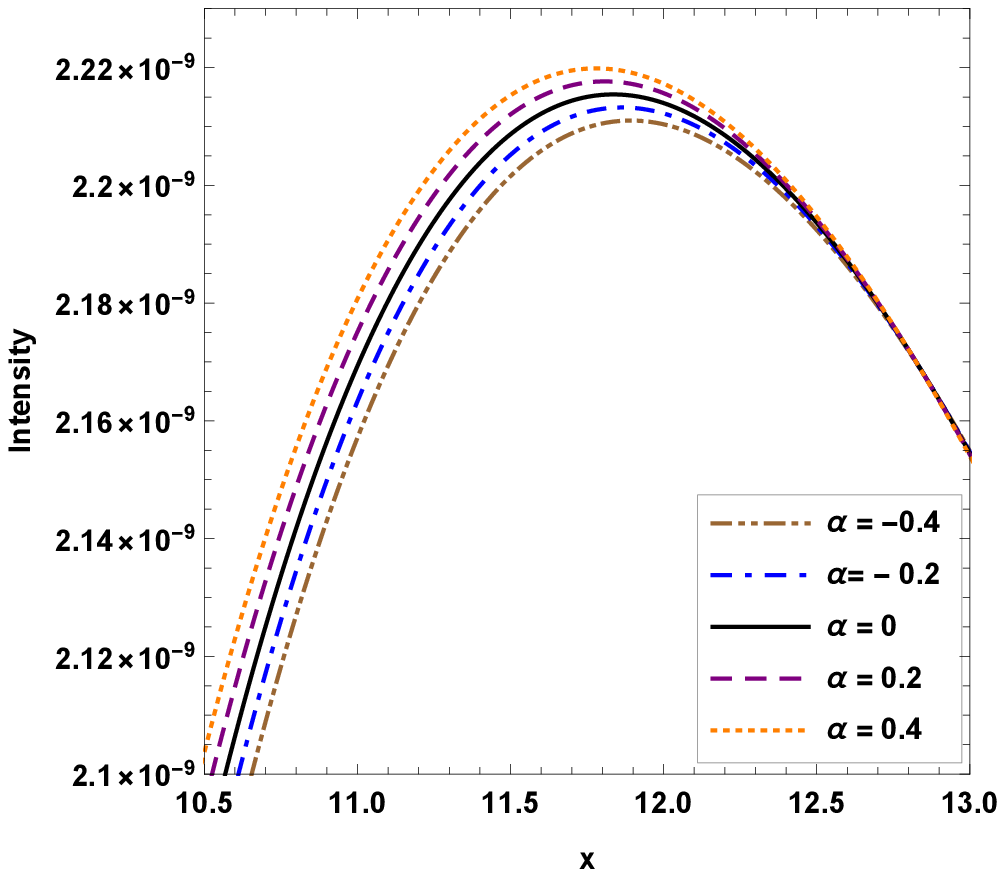}
\includegraphics[width=4.0cm]{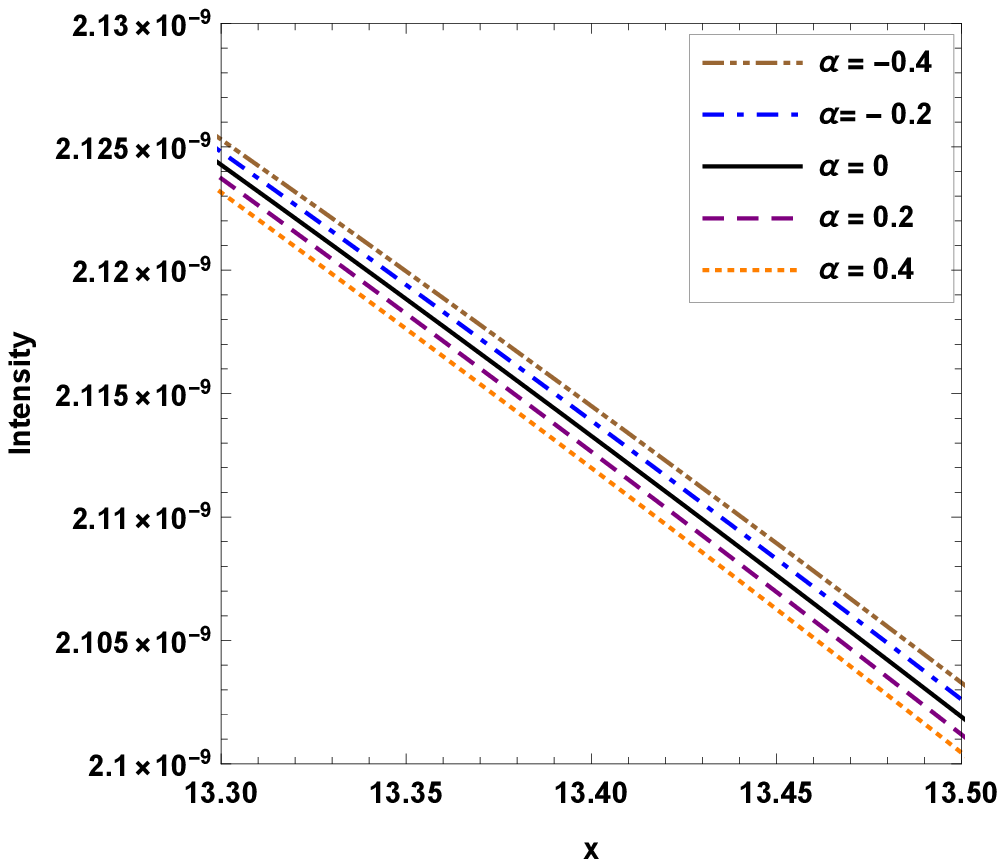}
\caption{The intensity distribution curve of images along the line $y=0$ for the observer with inclination angle $\theta_o=0^{\circ}$. The upper row is for the PPL and the bottle row is for the PPM, respectively. Here, we set $M=1$.}
\label{f4}
\end{figure}

\begin{figure}
\includegraphics[width=5.2cm]{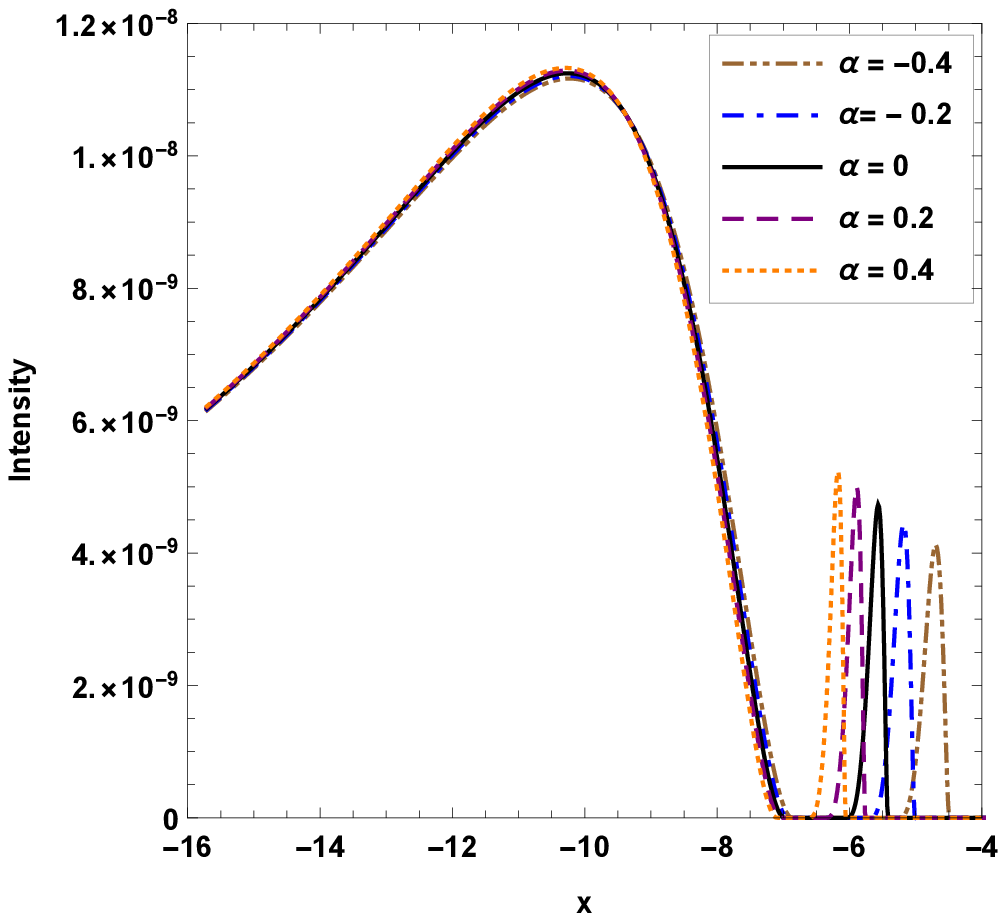}\;\;
\includegraphics[width=5.5cm]{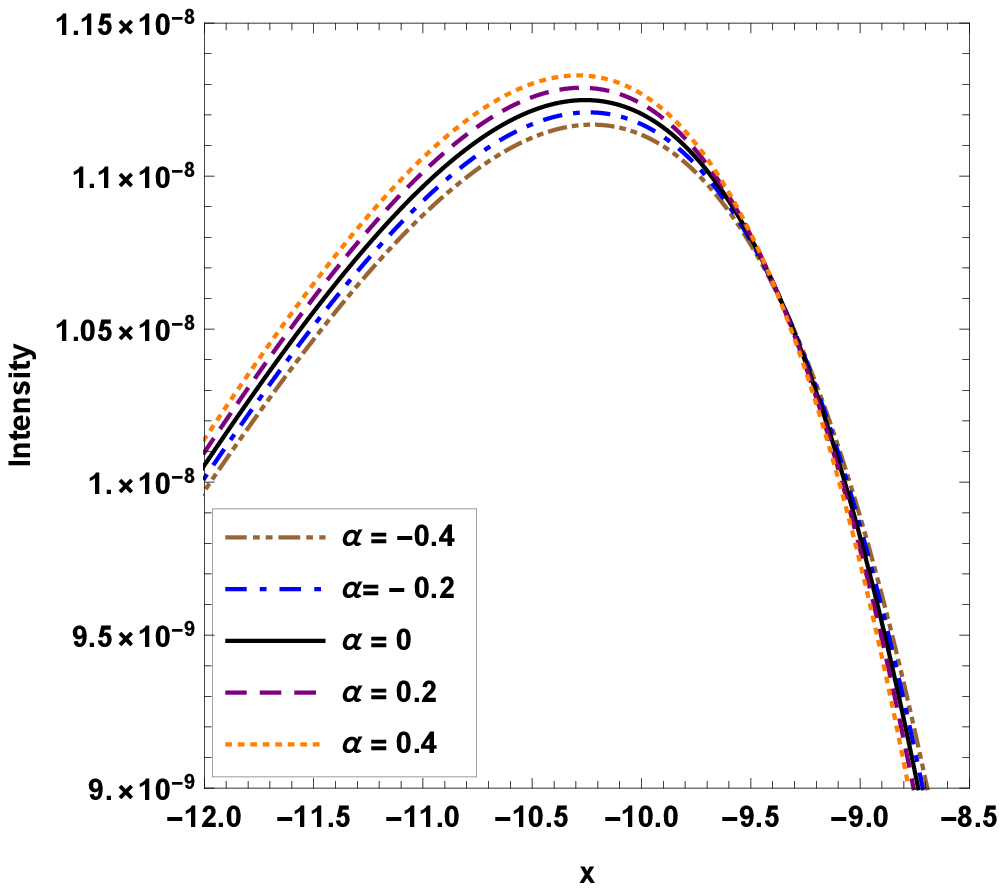}\;\;
\includegraphics[width=5.2cm]{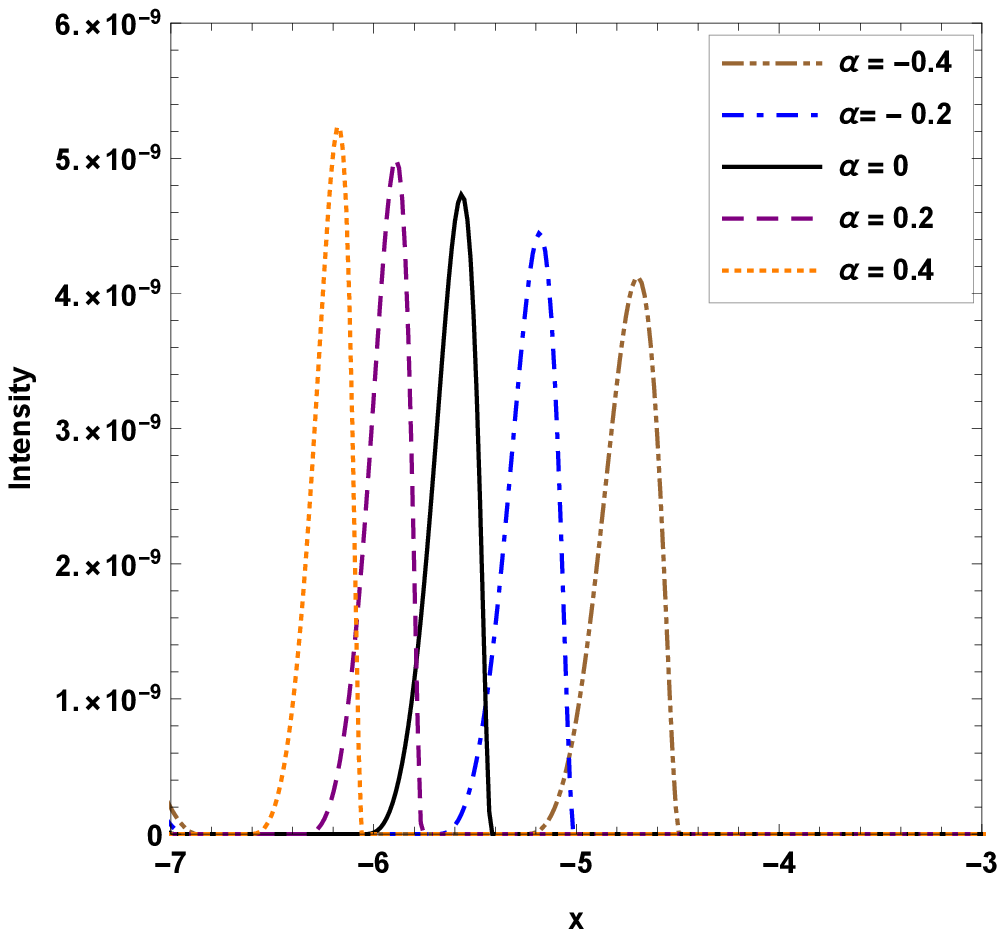}\\
\includegraphics[width=5.2cm]{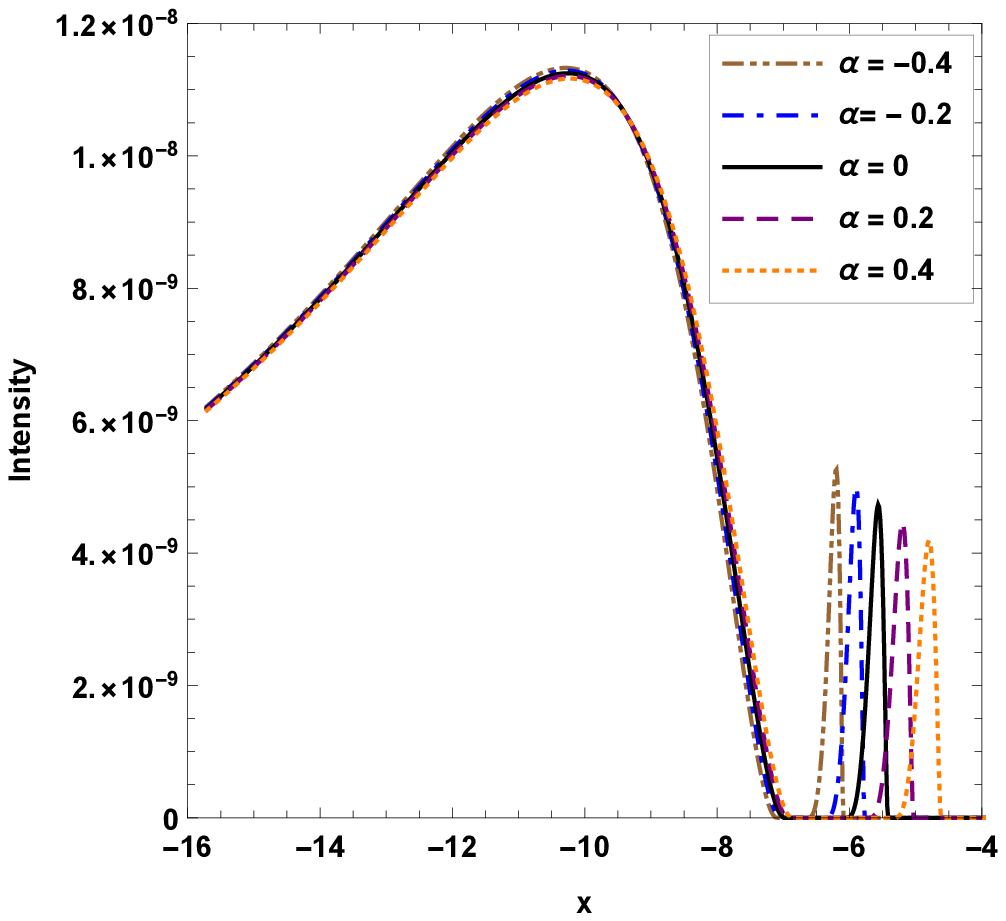}\;\;
\includegraphics[width=5.2cm]{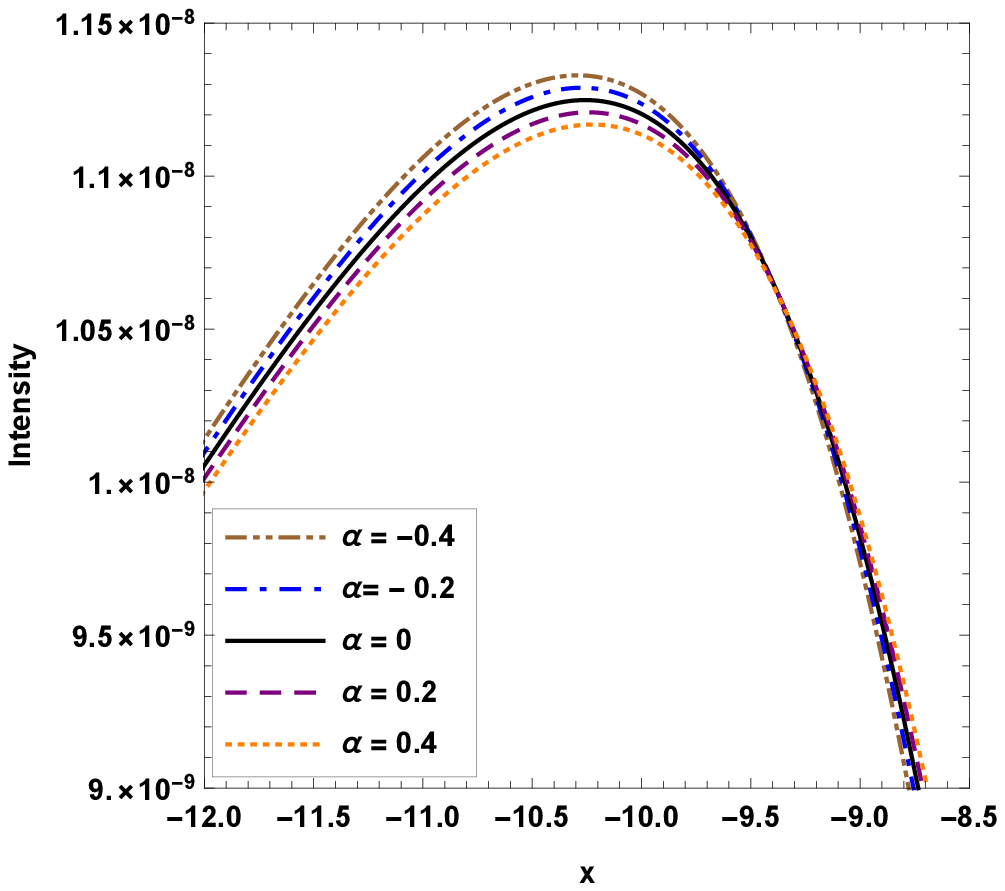}\;\;
\includegraphics[width=5.2cm]{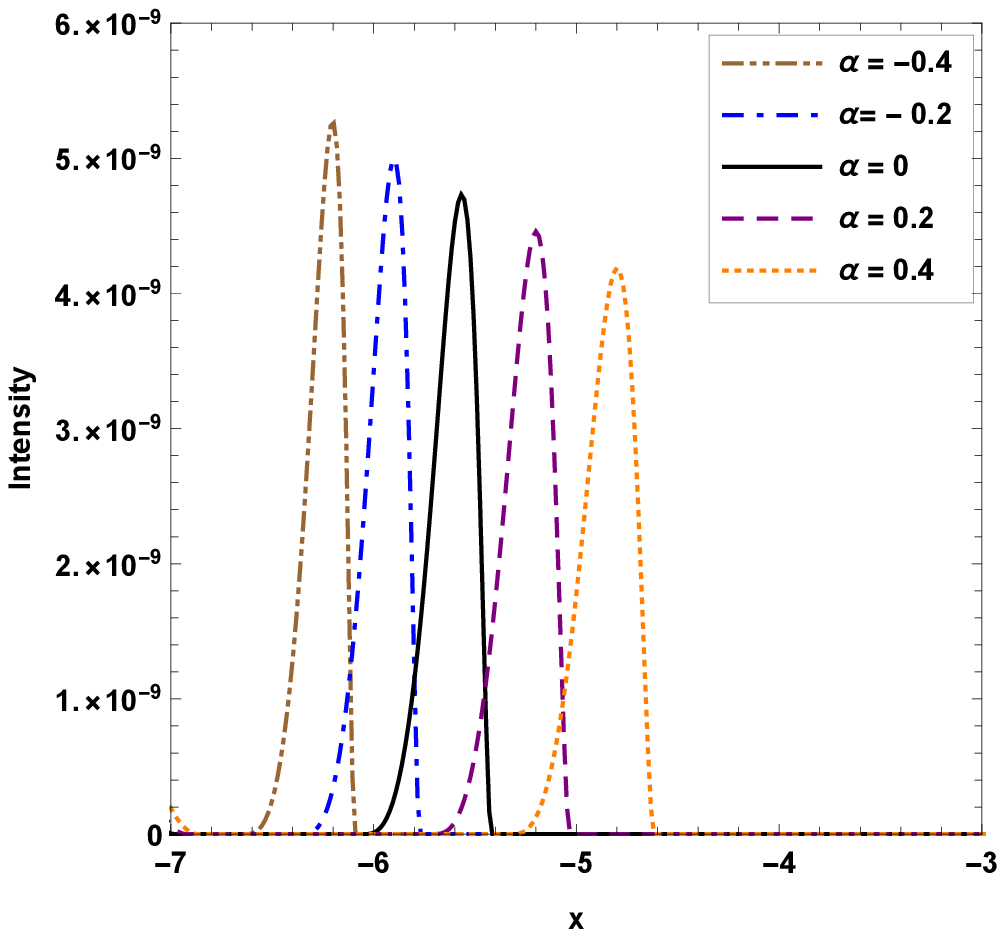}
\caption{The intensity distribution curve of images along the line $y=0$ for the observer with inclination angle $\theta_o=70^{\circ}$. The upper row is for the PPL and the bottle row is for the PPM, respectively. Here, we set $M=1$.}
\label{f5}
\end{figure}

In this section, we make use of the general relativistic ray-tracing code GYOTO \cite{gyoto} to
present the image of a Schwarzschild black hole with a thin disk caused by the polarized light coupled with Weyl tensor. Here, we assume that the disk of emitting matter around Schwarzschild black hole is a geometrically thin infinite accretion disk and is located in the  equatorial plane. In Figs. (\ref{f1})-(\ref{f3}), we present the image of a Schwarzschild black hole with a thin disk caused by PPL and PPM for the observer with different inclination angles. It is shown that the image size of photon ring increases with the coupling parameter $\alpha$ for the PPL, but decreases for the PPM.  Moreover, with the increasing of $\alpha$, the bright region in the image with the inclination angle  $\theta_o=70^{\circ}$ caused by the PPL extends to both sides along the black boundary and the disk's image in the high latitude zone shrinks. In the image caused by the PPM, one can find that the bright region shrinks along the black boundary and the size of the disk's image in the high latitude zone increases, which is just on the contrary to the PPL case.
In Figs.(\ref{f4}) and  (\ref{f5}), we also present the intensity distribution curve of image along the line $y=0$. For the direct image caused by the PPL,  we find that the intensity in the region near the black hole decreases with $\alpha$, but in the far region it increases. The change of the direct image's intensity with $\alpha$ for the PPM is the opposite of that for the PPL.
For the secondary image, its maximum intensity depends on the inclination angle  $\theta_o$ of the observer. As $\theta_o=0^{\circ}$, the maximum intensity of the secondary image is almost independent of the coupling parameter $\alpha$. However, as
$\theta_o=70^{\circ}$, it is an increasing function of $\alpha$ for the PPL and a decreasing function for the PPM. Moreover, we find that the width of the secondary image decreases with $\alpha$ for the PPL and increases for the PPM.

We are now to present the polarized image of a Schwarzschild black hole with a thin disk arising from  the coupling between photon and  Weyl tensor. Following the operation in refs.\cite{poimag3, poth1,poth2,poth3,pokerr}, in a Schwarzschild black hole spacetime, one can find that the Penrose-Walker constant $\kappa$ \cite{wpen} for a photon moving along a trajectory can be computed from its initial polarization $f^{\mu}$ and momentum $p^{\mu}$ at the source in the disk where $r_s$ and $\theta_s=\frac{\pi}{2}$,
\begin{figure}
\includegraphics[width=16cm]{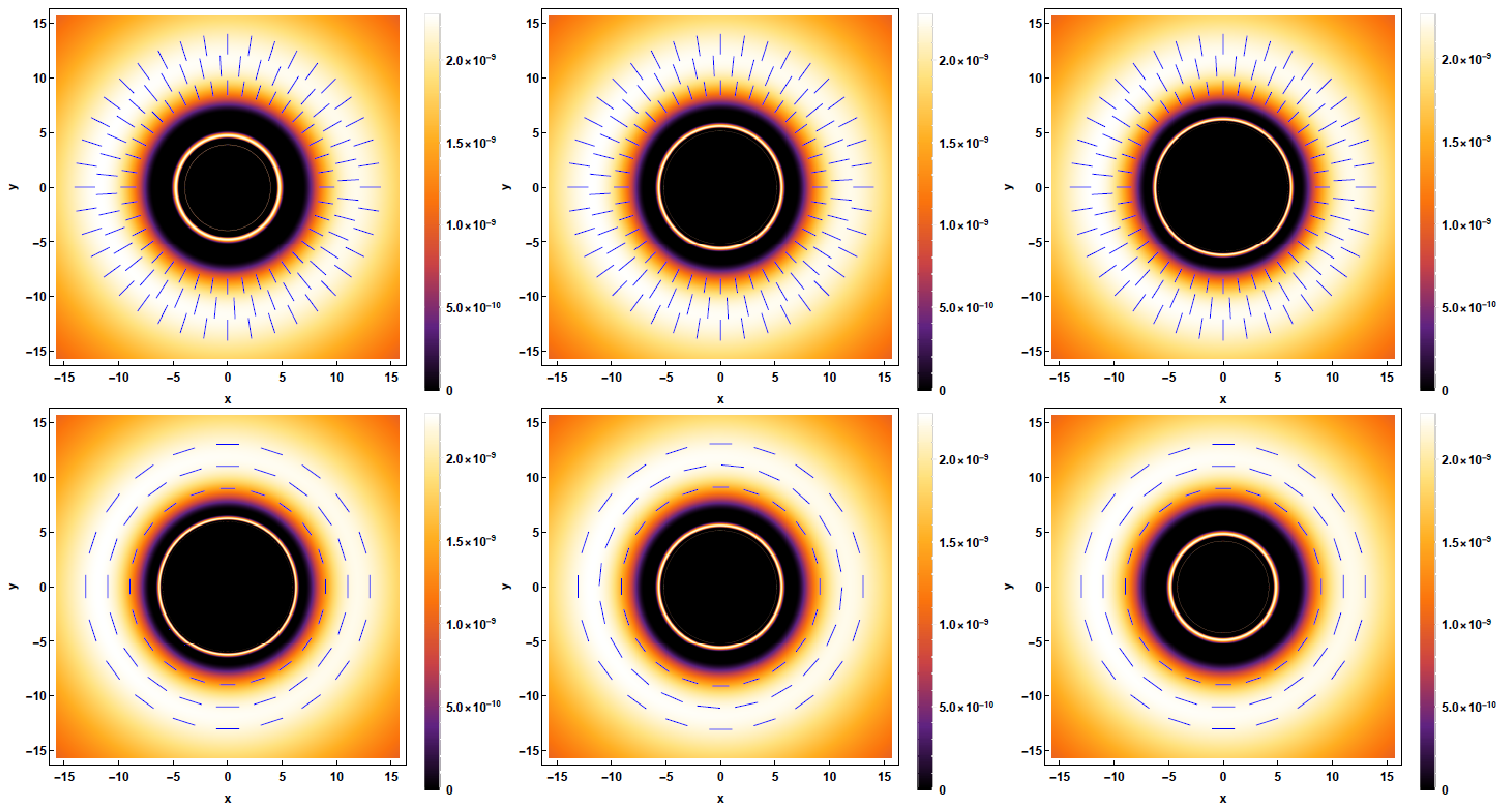}
\caption{Polarized intensity tick plots  in the direct image for the fixed $\theta_o=0^{\circ}$. The upper row is for the PPL and the bottom row is for the PPM.  In each row, the coupling parameter $\alpha$ in three panels from left to right is set to $-0.4$, $0$, and $0.4$, respectively. Here, we set $M=1$.}
\label{f60}
\end{figure}
\begin{figure}
\includegraphics[width=16cm]{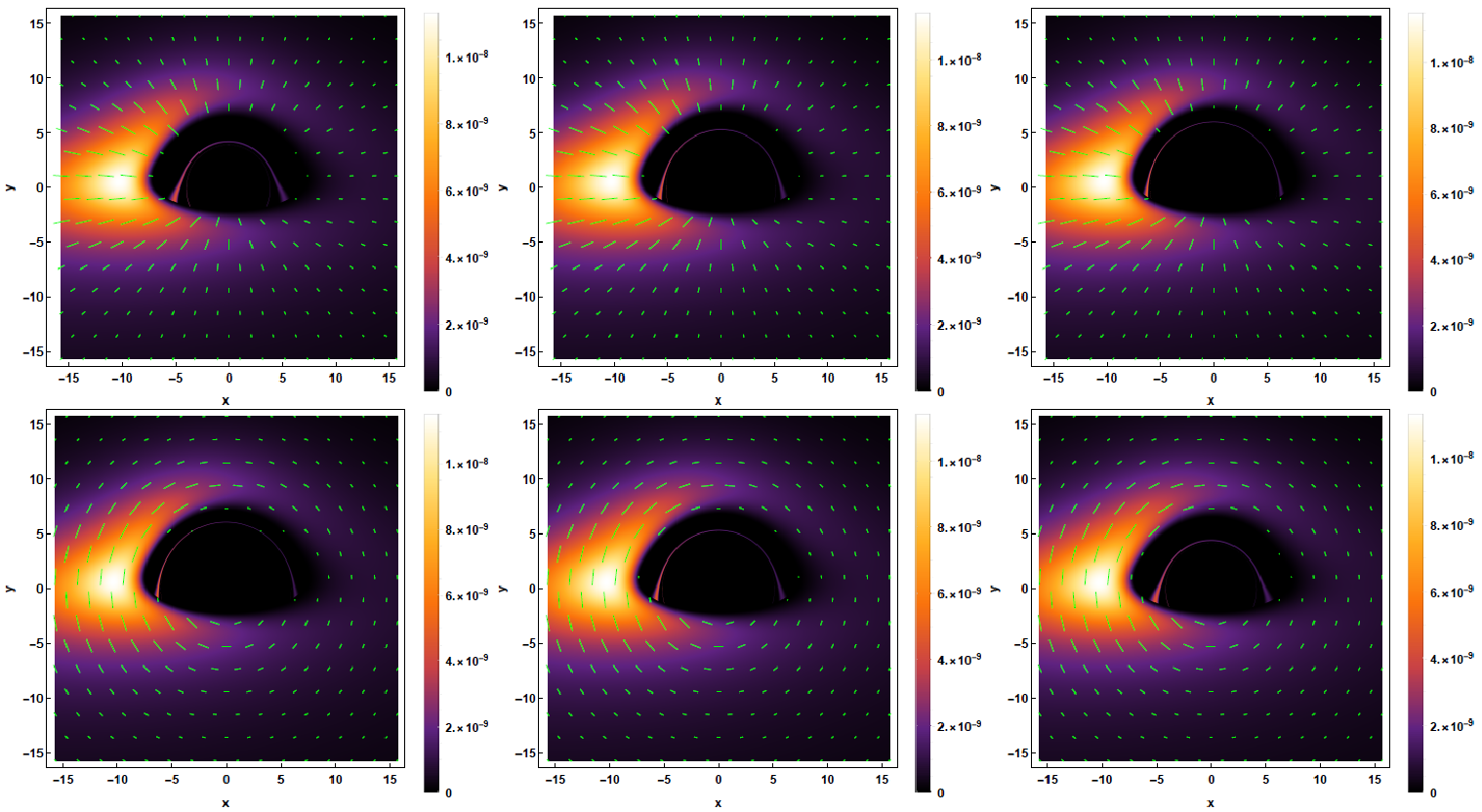}
\caption{Polarized intensity tick plots in the direct image for the fixed $\theta_o=70^{\circ}$. The upper row is for the PPL and the bottom row is for the PPM.  In each row, the coupling parameter $\alpha$ in three panels  from left to right is set to $-0.4$, $0$, and $0.4$, respectively. Here, we set $M=1$.}
\label{f602}
\end{figure}
\begin{eqnarray}
\kappa=\kappa_1+i\kappa_2=r_s(\mathcal{C}-i\mathcal{ D})
\end{eqnarray}
with
\begin{eqnarray}
\mathcal{C}=p^{t}f^{r}-p^{r}f^{t},\quad\quad\quad \mathcal{ D}=r^2_s(p^{\phi}f^{\theta}-p^{\theta}f^{\phi}).
\end{eqnarray}
And then the unit-normalized observed polarization $(f^{x}, f^{y})$ in the observer screen at position ($x,\;y)$ can be expressed as
\begin{eqnarray}
\vec{f}=(f^{x},f^{y})=\bigg(\frac{y\kappa_2+x\kappa_1}{\sqrt{(\kappa^2_1+\kappa^2_2)(x^2+y^2)}},
\frac{y\kappa_1-x\kappa_2}{\sqrt{(\kappa^2_1+\kappa^2_2)(x^2+y^2)}}\bigg).
\end{eqnarray}
For the  polarization vector $l^{\mu}=(k^{r},k^{t},0,0)$, we have $\kappa_2=-r_s\mathcal{ D}=0$. And then,
its unit-normalized observed polarization $(f^{x}_l, f^{y}_l)$ can be expressed as
\begin{eqnarray}
\vec{f}_l=(f^{x}_l,f^{y}_l)=(\frac{x}{\sqrt{x^2+y^2}},
\frac{y}{\sqrt{x^2+y^2}}).
\end{eqnarray}
This means that the observed polarization $\vec{f}_l$ for $l^{\mu}$ is along the ``radial '' direction $\vec{r}=x\vec{i}+y\vec{j}$  in the sky plane. Similarly, for the  polarization vector $m^{\mu}=(0,0,-k^{\phi},k^{\theta})$, we have $\kappa_1=r_s\mathcal{C}=0$ and its corresponding unit-normalized observed polarization $(f^{x}_m, f^{y}_m)$ is
\begin{eqnarray}
\vec{f}_m=(f^{x}_m,f^{y}_m)=(\frac{y}{\sqrt{x^2+y^2}},
\frac{-x}{\sqrt{x^2+y^2}}).
\end{eqnarray}
And then the observed polarization $\vec{f}_m$  is along the ``angular '' direction in the sky plane. It is
obvious that the observed polarization vectors $\vec{f}_l$ and $\vec{f}_m$ are perpendicular to each other, which is also shown in Figs. (\ref{f60}) and (\ref{f602}).
Thus,  for a certain linear polarization light, one can compute
its total observed polarization $(f^{x}, f^{y})$ by its observed intensity components $I_l$ and $I_m$  along the vectors  $\vec{f}_l$ and $\vec{f}_m$ in the sky plane, i.e.,
\begin{eqnarray}
f^{x}=\sqrt{I_l}\cos\gamma-\sqrt{I_m}\sin\gamma,\quad\quad\quad f^{y}=\sqrt{I_l}\sin\gamma+\sqrt{I_m}\cos\gamma,
\end{eqnarray}
where $\gamma\equiv \arctan(\frac{y}{x})$.
\begin{figure}
\includegraphics[width=16cm]{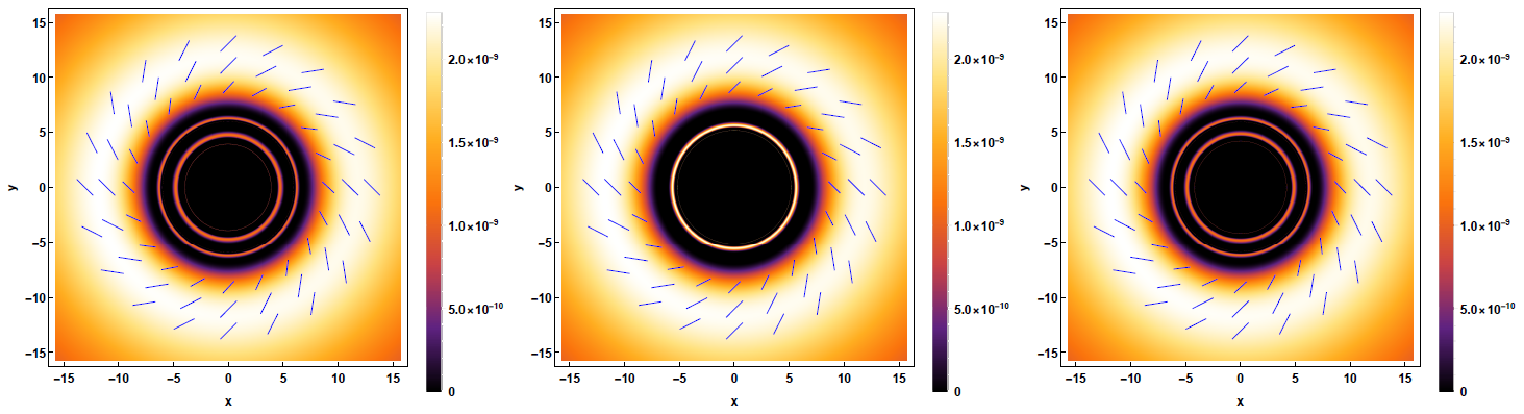}
\caption{Polarized intensity tick plots in the direct image for the fixed $\theta_o=0^{\circ}$. The coupling parameter $\alpha$ in three panels  from left to right is set to $-0.4$, $0$, and $0.4$, respectively. Here, we set $M=1$.}
\label{f8}
\end{figure}
\begin{figure}
\includegraphics[width=16cm]{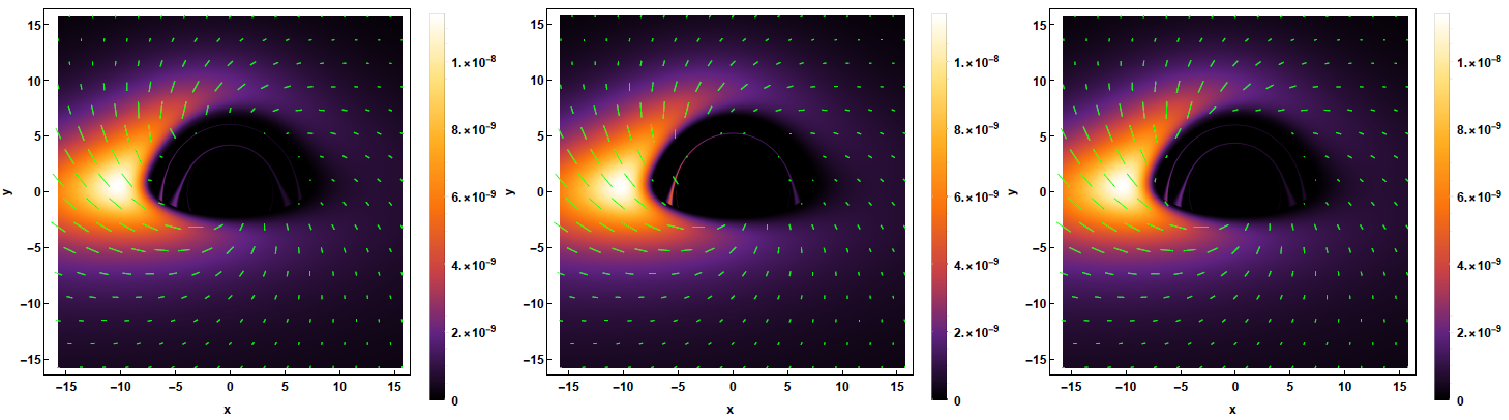}
\caption{Polarized intensity tick plots in the image for the fixed $\theta_o=70^{\circ}$. The coupling parameter $\alpha$ in three panels  from left to right is set to $-0.4$, $0$, and $0.4$, respectively. Here, we set $M=1$.}
\label{f9}
\end{figure}
In Figs.(\ref{f8}) and (\ref{f9}), we show the polarized pattern of an equatorial thin disk around a Schwarzschild black hole in the sky plane for different coupling parameter $\alpha$. As $\theta_o=0^{\circ}$, we find that the polarized intensity tick plot has a counterclockwise vortex-like distribution with a rotational symmetry. As  $\theta_o=70^{\circ}$, the rotational symmetry vanishes in the corresponding tick plot.
\begin{figure}
\includegraphics[width=5.5cm]{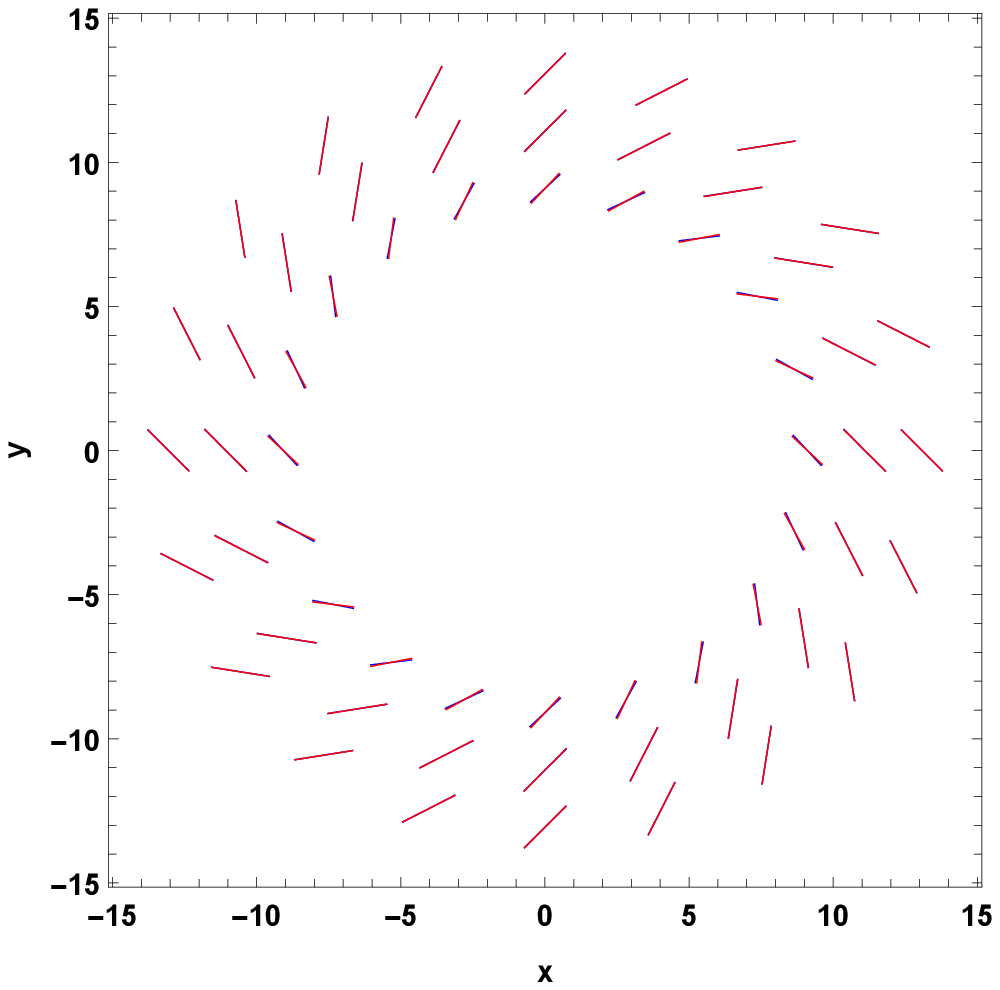}
\includegraphics[width=5.5cm]{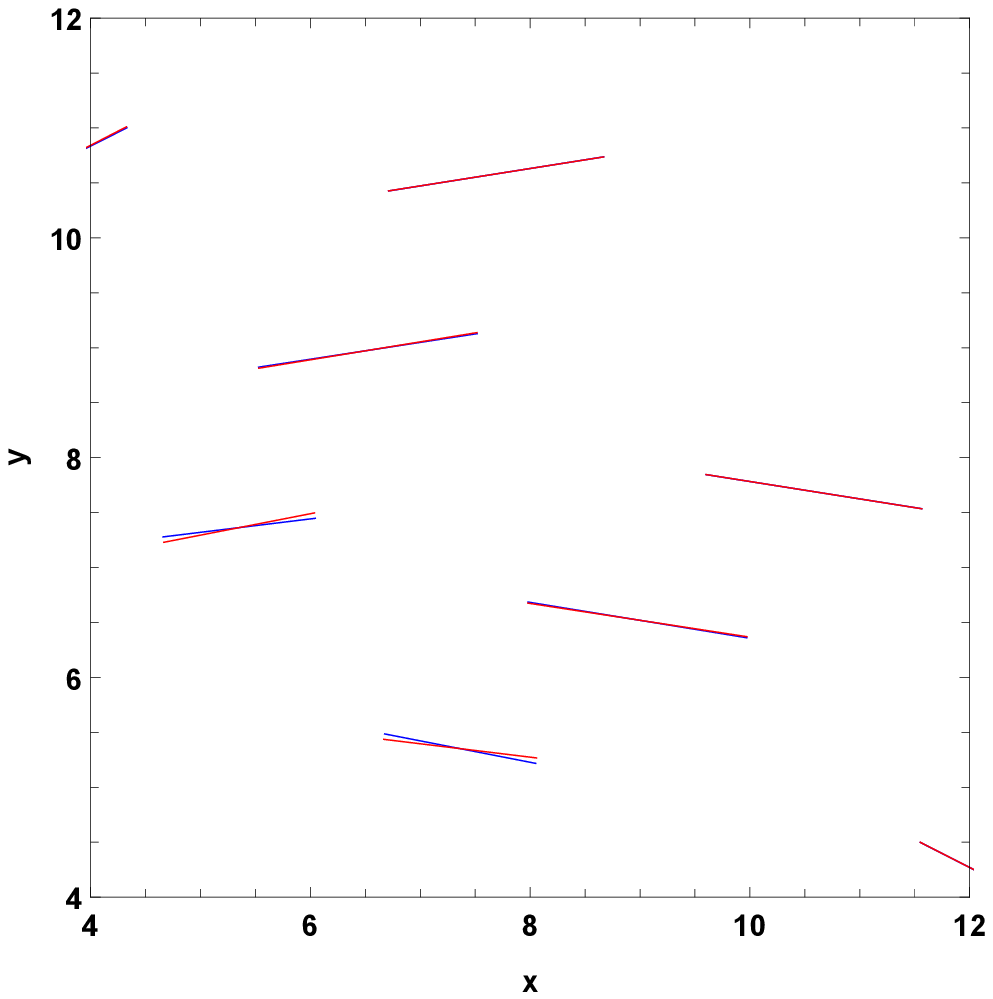}
\includegraphics[width=5.4cm]{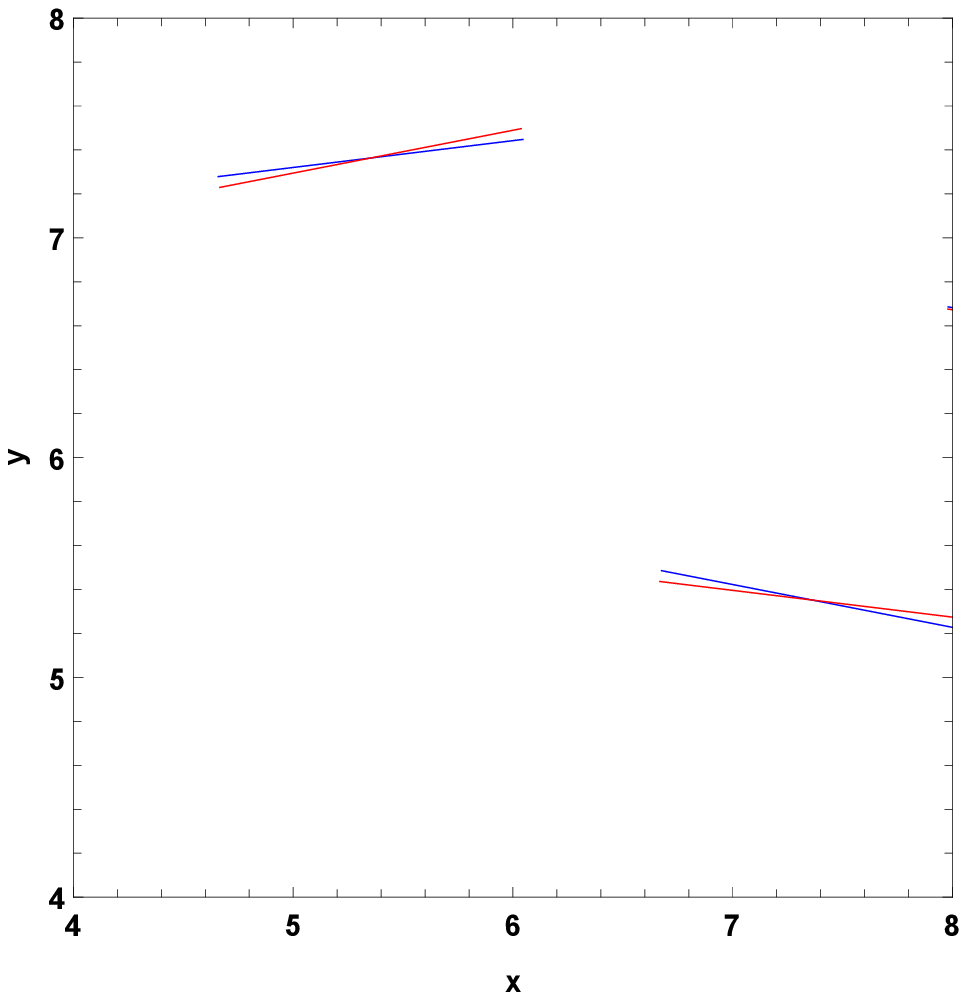}
\caption{Difference of the total observed polarization in the cases $\alpha=-0.4$ and $\alpha=0.4$ for $\theta_o=0^{\circ}$. The red line and the blue line correspond to $\alpha=-0.4$ and $0.4$, respectively. The panel on the right is an enlargement of that in the left. Here, we set $M=1$.}
\label{f10}
\end{figure}
\begin{figure}
\includegraphics[width=5.5cm]{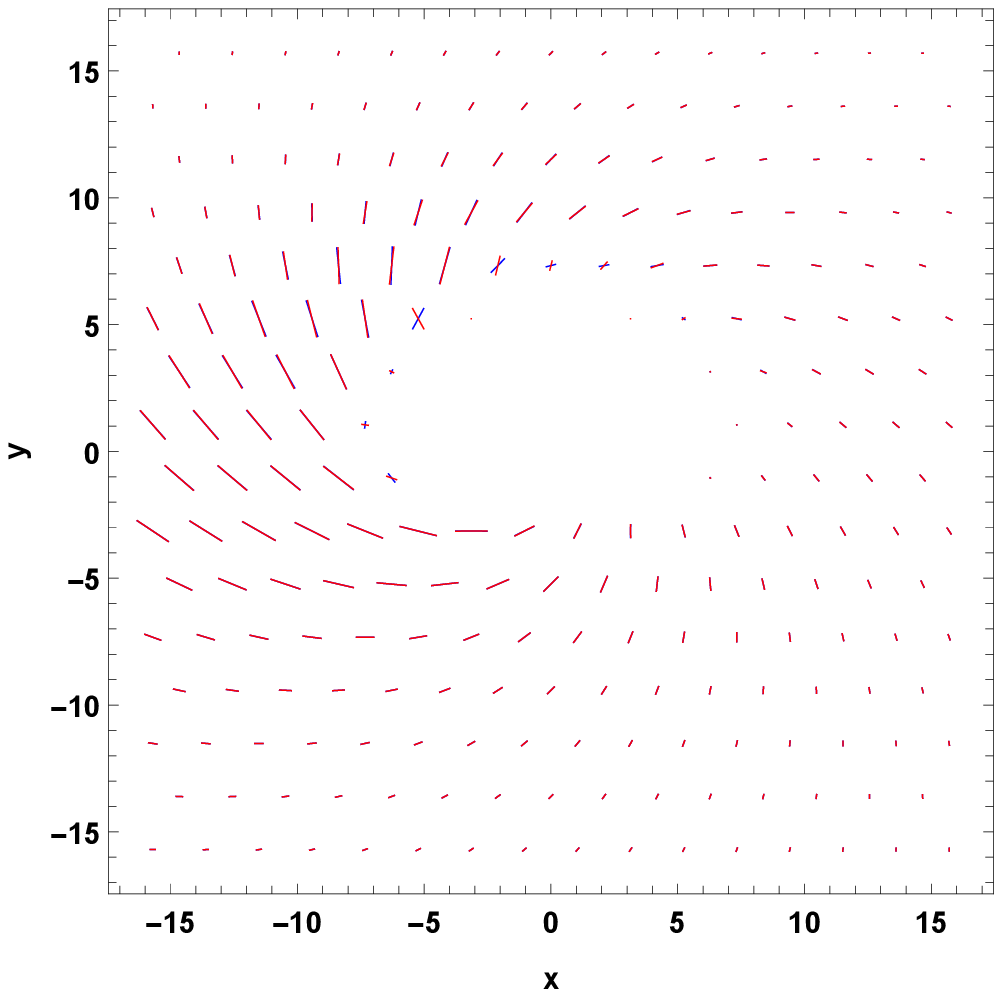}
\includegraphics[width=5.4cm]{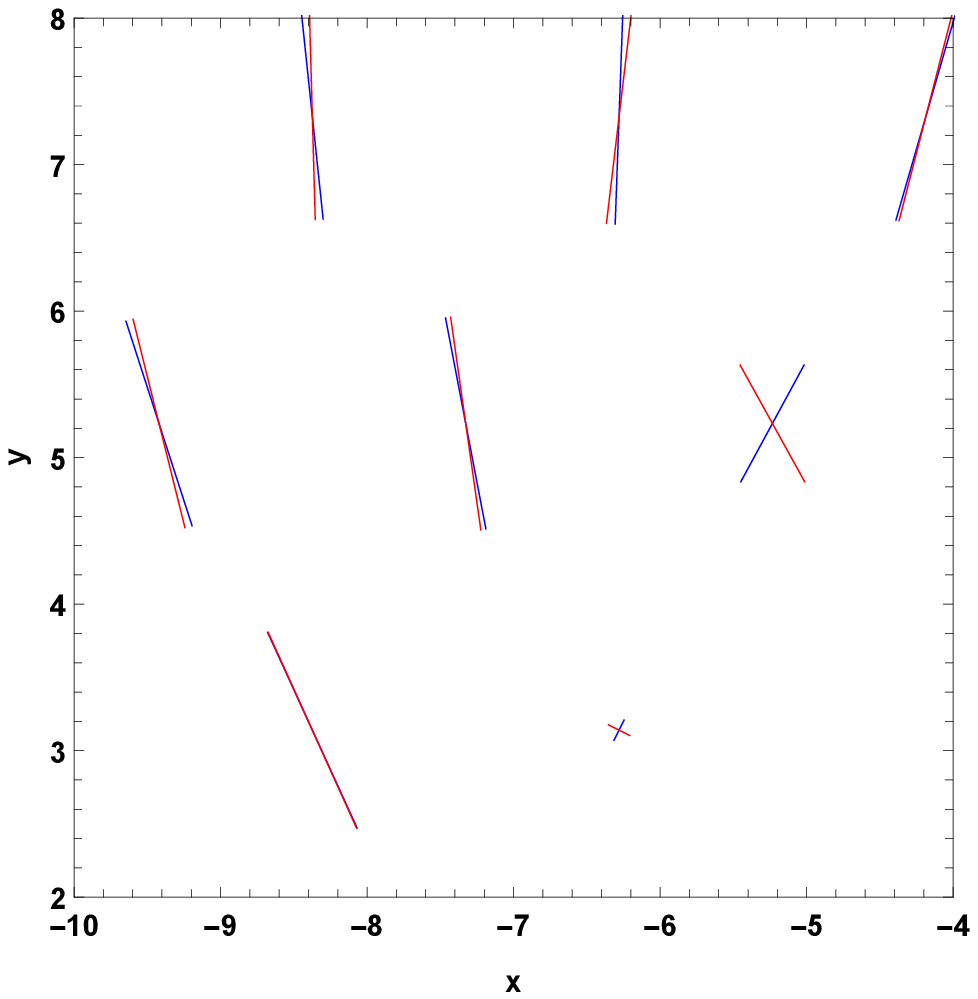}
\includegraphics[width=5.4cm]{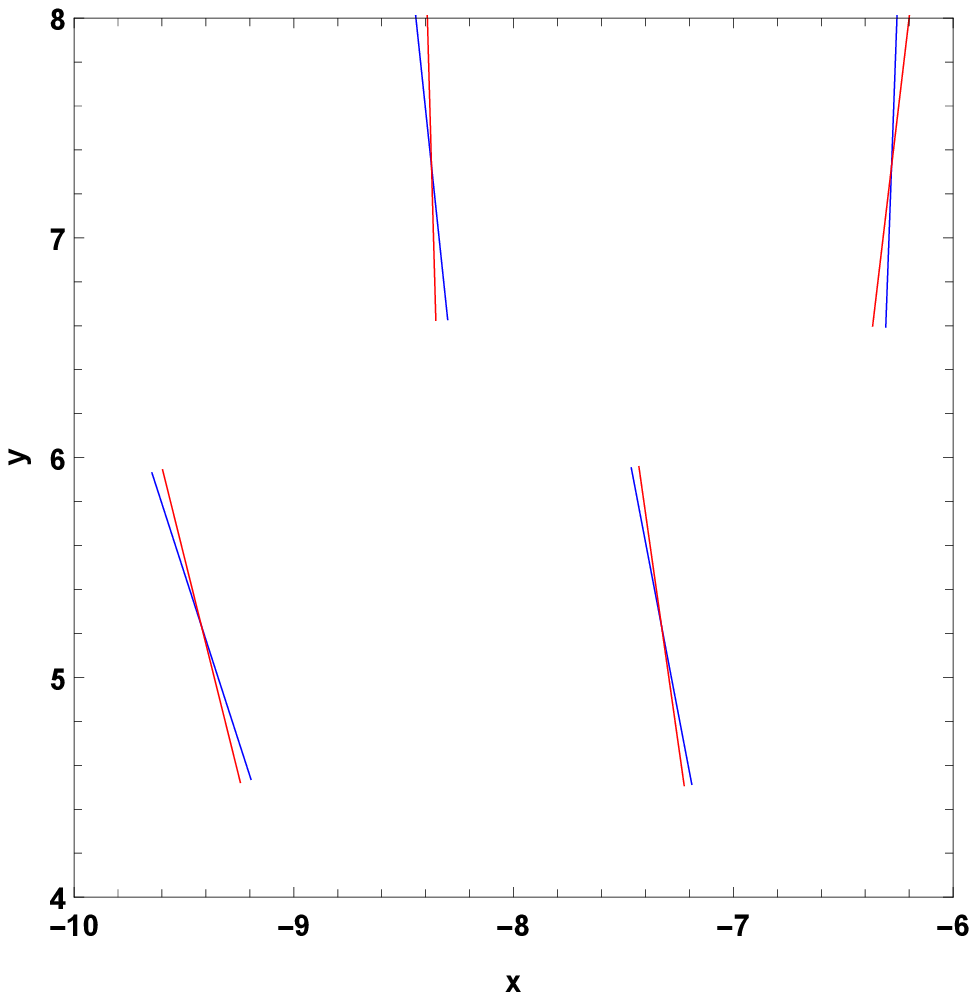}
\caption{Difference of the total observed polarization in the cases $\alpha=-0.4$ and $\alpha=0.4$ for $\theta_o=70^{\circ}$. The red line and the blue line correspond to $\alpha=-0.4$ and $0.4$, respectively. The panel on the right is an enlargement of that in the left. Here, we set $M=1$.}
\label{f11}
\end{figure}
Moreover, we find that the observed polarized intensity in the bright region is stronger than that in the darker region. Figs.(\ref{f8})-(\ref{f11}) also show that the effect of $\alpha$ on the observed polarized vector is weak in general and the  stronger effect of $\alpha$ appears in the bright region close to black hole in the image plane.

Finally, we discuss briefly the constraint on the coupling constant $\alpha$ related to the interaction between  Maxwell field and Weyl tensor. From the deflection of light near the Solar, one find that the upper bound of the coupling constant is $|\alpha|\leq4.3\times10^{13} m^2$ \cite{limitalpha}. In Figs.(\ref{f1})-(\ref{f11}), we take the maximum absolute value of $\alpha$ is $|\alpha|=0.4 G^2M^2_{BH}/c^4$. This means that for the black holes with mass $M_{BH}=\frac{c^2}{G}\sqrt{\frac{|\alpha|}{0.4}}\leq7031M_{\odot}$, the values of $\alpha$ in Figs.(\ref{f1})-(\ref{f11})  satisfy the constraint on the coupling constant from the Solar System.

\section{Summary}

We have studied polarized image of a Schwarzschild black hole with a thin accretion disk produced by photon coupled to Weyl tensor. Our results show that the black hole shadow, the thin disk pattern and the observed polarized vector depend on the coupling between photon and Weyl tensor. With the coupling parameter $\alpha$, the image size of photon ring increases for the PPL and
decreases for the PPM.  For the direct image caused by the PPL,  its intensity decreases in the region near the black hole,  but it increases in the far region. The change of the direct image's intensity with $\alpha$ for the PPM is the opposite of that for the PPL. For the secondary image, its maximum intensity depends on the inclination angle  $\theta_o$ of the observer. As $\theta_o=0^{\circ}$, the maximum intensity of the secondary image is almost independent of the coupling parameter $\alpha$. As
$\theta_o=70^{\circ}$, it is an increasing function of $\alpha$ for the PPL and a decreasing function for the PPM. The width of the secondary image decreases with $\alpha$ for the PPL and increases for the PPM. Moreover, the bright region in the image with the inclination angle  $\theta_o=70^{\circ}$ caused by the PPL extends to both sides along the black boundary and the disk's image in the high latitude zone shrinks. In the image caused by the PPM, one can find that the bright region shrinks along the black boundary and the size of the disk's image in the high latitude zone increases, which is just on the contrary to the PPL case.

We also present the tick plot for the observed polarized intensity. As $\theta_o=0^{\circ}$, we find that the polarized intensity tick plot has a counterclockwise vortex-like distribution with a rotational symmetry. As  $\theta_o=70^{\circ}$, the rotational symmetry vanishes in the corresponding tick plot.
Moreover, we find that the observed polarized intensity in the bright region is stronger than that in the darker region. It is also noted that the effect of $\alpha$ on the observed polarized vector is weak in general and the  stronger effect of $\alpha$ appears in the bright region close to black hole in the image plane. These features in the polarized image
could help us to understand black hole shadow, thin accretion disk and the coupling between photon and Weyl tensor.

\section{\bf Acknowledgments}
This work was  supported by the National Natural Science
Foundation of China under Grant No.12035005, 11875026, 11875025 and 2020YFC2201403.

\end{document}